# Exploring the energy landscape of resistive switching in antiferromagnetic $Sr_3Ir_2O_7$

Morgan Williamson[1,2], Shida Shen[1,2], Gang Cao[3], Jianshi Zhou[2], John B. Goodenough[2], Maxim Tsoi[1,2]

[1]Physics Department, The University of Texas at Austin, Austin, Texas 78712, USA

[2]Texas Materials Institute, The University of Texas at Austin, Austin, Texas 78712, USA

[3]Department of Physics, University of Colorado-Boulder, Boulder, CO 80309, USA

## ABSTRACT

We study the resistive switching triggered by an applied electrical bias in antiferromagnetic Mott insulator $Sr_3Ir_2O_7$. The switching was previously associated with an electric-field driven structural transition. Here we use time-resolved measurements of the switching to probe the energy barrier associated with the transition. We quantify the changes in the energy barrier height with respect to the applied bias and find a linear decrease of the barrier with increasing bias. Our observations support the potential of antiferromagnetic transition metal oxides for spintronic applications.

Antiferromagnetic (AFM) materials are expected to improve stability, scalability, and speed of spintronic applications, e.g. magnetic memories, thanks to the insensitivity of AFMs to magnetic fields and their high natural frequencies. Of particular interest are AFM transition metal oxides (TMO) as their properties can be tuned using various external stimuli, thus opening an entirely new dimension to the field of spintronics. Recently, we have demonstrated that the transport properties of AFM TMOs – Mott insulators $Sr_2IrO_4$ and $Sr_3Ir_2O_7$ – can be tuned by an externally applied electric field [1, 2]. For instance, we found a reversible resistive switching driven by an applied electric bias [1, 2]. The switching was tentatively attributed to electric-field driven lattice distortions/structural transition with potential applications for writing in AFM memory applications.

In this Letter we probe the energy barrier associated with this transition using time-resolved measurements of the switching. We observe an exponential dependence of the switching probability on both electric bias and temperature consistent with thermal activation over an energy barrier. We quantify the changes in the energy barrier height with respect to the applied bias and find a linear decrease of the barrier with increasing bias. Our observations elucidate the activation picture of resistive switching in Mott insulators and support the potential of antiferromagnetic transition metal oxides for spintronic applications.

Single crystals of $Sr_3Ir_2O_7$ were synthesized for this work using a self-flux technique described elsewhere [3]. The technique results in single crystalline flakes of $Sr_3Ir_2O_7$ with thicknesses of about 0.2-mm and the [001] c-axis oriented perpendicular to the flake's surface (area ~0.5×0.5 $mm^2$). Single crystal X-ray diffraction was employed to determine the crystal structure of the flakes. Ag paste and In metal contacts (area ~0.2×0.2 $mm^2$) on top and bottom surfaces of $Sr_3Ir_2O_7$ flakes were used to supply and sink electrical currents (up to 15 mA). In such a 2-probe geometry the applied currents flow (primarily) along the [001] c-axis of the crystal and were shown previously [4] to trigger a resistive switching at sufficiently high electrical bias. The switching was monitored by transport measurements – current-voltage (*I-V*) characteristics of the crystal were measured with the sample in liquid nitrogen (LN) in the temperature range from 77-80 K. The temperature control in this range was achieved by varying the LN pressure from 750-1020 Torr using an adjustable pressure relief valve on the LN cryostat. Time domain measurements were enabled by a Teledyne LeCroy Wavesurfer 3034 oscilloscope (350 MHz bandwidth, 4 GS/s) which was used to measure a delay time between the application of a bias current and the resistive switching. The delay time is projected to be conceptually similar to the decay time associated with the probabilistic processes governing radioactive decay.

Figure 1 shows a representative current-voltage (*I-V*) characteristic of $Sr_3Ir_2O_7$ measured at 77 K – the crystal resistance $R=V/I$ is shown as a function of the bias current *I*. Here the black and grey curves show the *R(I)* for up- and down-sweeps of *I*, respectively. For an increasing bias, both positive and negative, we observe a continuous decrease of the crystal resistance and two irreversible switching events to a higher resistance (indicated by arrows in Fig. 1) at critical biases $I_C \approx \pm 8.5$ mA and $\pm 12.7$ mA. For a decreasing bias, only one switching back to a lower resistance is observed at around $\pm 5$ mA. Similar switching characteristics were previously observed in both $Sr_2IrO_4$ and $Sr_3Ir_2O_7$ [2, 4] where the continuous/irreversible changes in resistance were associated with electric-field driven lattice distortions/structural transition. The inset to Fig. 1 shows the temperature dependence of the critical current associated with the lower-bias switching event at around 8.5 mA. Increasing the temperature of the sample has the effect of decreasing the critical bias current with a linear dependence quantified by $119 \pm 3$ µA/K.

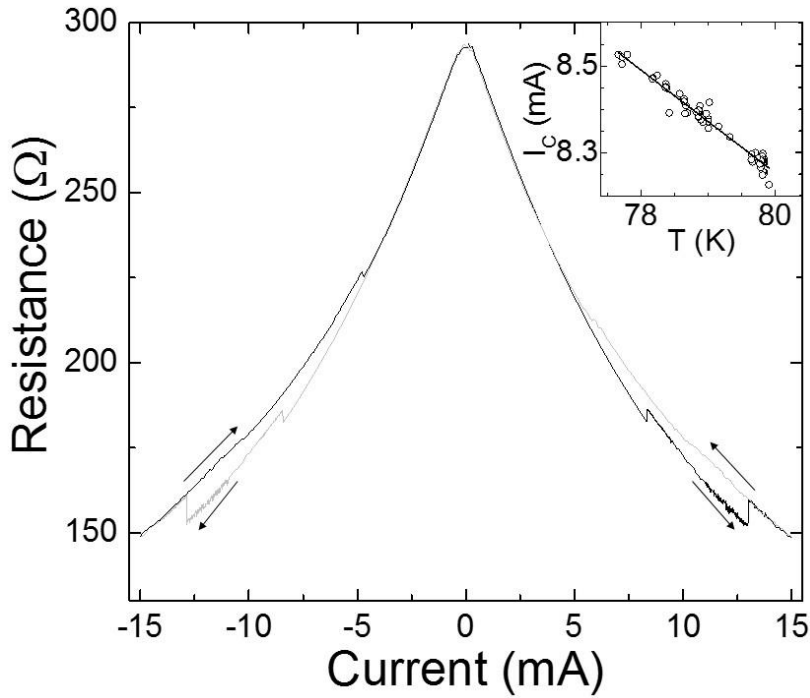

**Figure 1:** (a) Resistive switching in *R(I)* characteristic of $Sr_3Ir_2O_7$ (T = 78.6 K). Gray and black traces represent the up- and down-sweeps of *I*, respectively. Arrows indicate switching events. Two switching events to a higher resistance for increasing bias (at $I_c=\pm 8.5$ mA and $\pm 12.7$ mA) and one to a lower resistance for decreasing bias are clearly visible. The inset shows the temperature dependence of the critical current associated with the first switching event at positive bias.

Next we investigate the onset of the resistive switching in the time domain. We will focus on the lower-bias switching event around 8.5 mA (see Fig. 1). The procedure for triggering the switching consisted of slowly increasing the applied electrical bias from zero to a value (usually 8 mA) well below $I_C$ and then abruptly (~150 μs) increasing the applied bias to a value above $I_C$. This abrupt jump in bias naturally produces an increase in the voltage across the crystal which has been used to trigger the oscilloscope. Figure 2 shows the resulting time evolution of voltage $V_{OSC}$ across the sample measured by oscilloscope for a bias jump from 8 mA to 9.17 mA. Here the first rise in the oscilloscope voltage $V_{OSC}$ at 0 ms is associated with the abrupt bias jump from 8 to 9.17 mA. Furthermore we can clearly see a second rise in $V_{OSC}$ at about 15 ms which is produced by an increase in the sample resistance which occurs with a delay $\tau \approx 15$ ms after the bias jump. We thus conclude that the resistive switching requires a certain time $\tau$ at a given bias. We have repeated the above measurement 600 times and found a distribution of delay times shown in the inset to Fig. 2. The associated Poisson fit (solid curve in the inset) suggests that on average the switching at $I = 9.17$ mA is delayed by about 50 ms.

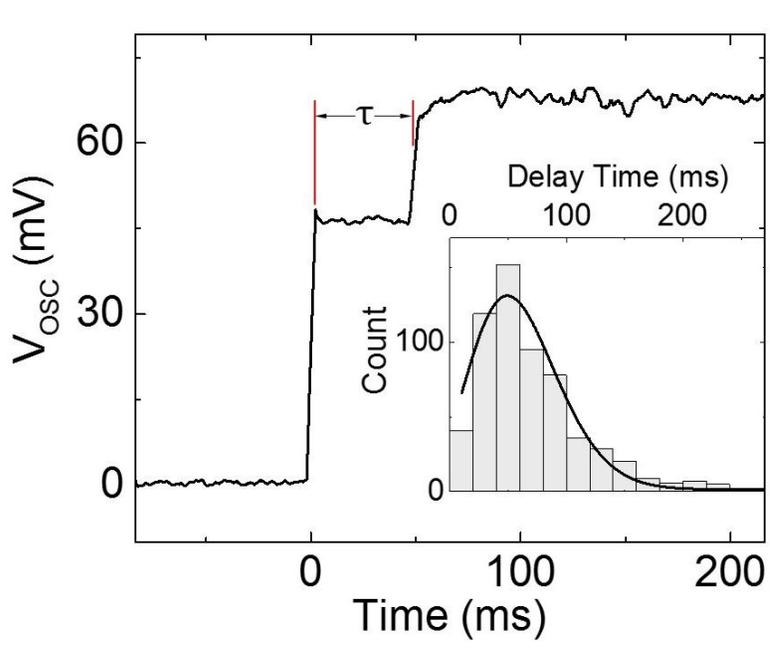

**Figure 2:** (a) A time trace of the process of switching from low bias to the '1st high-resistance' state. The vertical axis plots the voltage across the sample captured with an oscilloscope where the first abrupt increase is caused by a sudden rise in applied current. After a time, $\tau$, the switching occurs resulting in a second abrupt rise in measured voltage due to an increase in sample resistance indicating that switching has occurred. Figure 2b shows the results of the mentioned measurement repeated 600 times. A histogram corresponding to the relative frequency of measured delay time values is plotted with a Poisson fit.

We can use the measured delay time to determine the activation energy of switching. A simple model where the switching is associated with overcoming an energy barrier $\Delta$ implies the Arrhenius equation for the delay time:

$$\tau = \tau_0\, e^{\frac{\Delta}{k_B T}} \qquad [1]$$

with $\tau_0$ the attempt time (reciprocal of the attempt frequency), $k_B$ Boltzmann constant, and $T$ temperature. Assuming $\tau_0 = 1$ ns and using $T = 77$ K in our experiment, this analysis yields $\Delta = 50$ meV. While the exact value of $\tau_0$ in our system is unknown, this analysis provides a reasonable estimation of $\Delta$, since decreasing $\tau_0$ by as much as three orders of magnitude (to 1 ps) would increase $\Delta$ by only 20 meV. In contrast, the variation of the energy barrier $\Delta$ with respect to the applied bias current $I$ can be characterized exactly as we show next.

The procedure for finding the average delay time at a given switching current $I$ was repeated for different $I$s. These measurements have provided the $\tau(I)$ dependence displayed in Fig.3. Here open symbols show that the delay time increases significantly with decreasing bias current $I$. By applying Eq. 1 to every $\tau$ data point in Fig. 3 we have reconstructed the dependence of $\Delta$ on $I$. The inset to Fig. 3 shows that the energy barrier $\Delta$ decreases approximately linearly with the applied bias current $I$ (see the linear fit in Fig.3 inset). This result correlates with the field-effect model, which was used to explain the linear relationship between the activation energy and applied electrical bias observed in $Sr_3Ir_2O_7$ [4]. The field-effect model suggests that the switching barrier in Eq. 1 can be expressed as a function of the applied bias as:

$$\Delta = \Delta_0 - \gamma I \qquad [2]$$

where $\Delta_0$ is the barrier at zero $I$ and $\gamma$ is the parameter which characterizes the strength of the field effect. From the linear data fit in Fig. 3 inset we now define the $\gamma$ parameter – 172 eV/A. Using this parameter we have successfully fitted the $\tau(I)$ data in Fig. 3 (see solid curve) by an exponential function which combines Eqs. 1 and 2:

$$\tau = \tau_0\, e^{\frac{\Delta_0 - \gamma I}{k_B T}} \qquad [3]$$

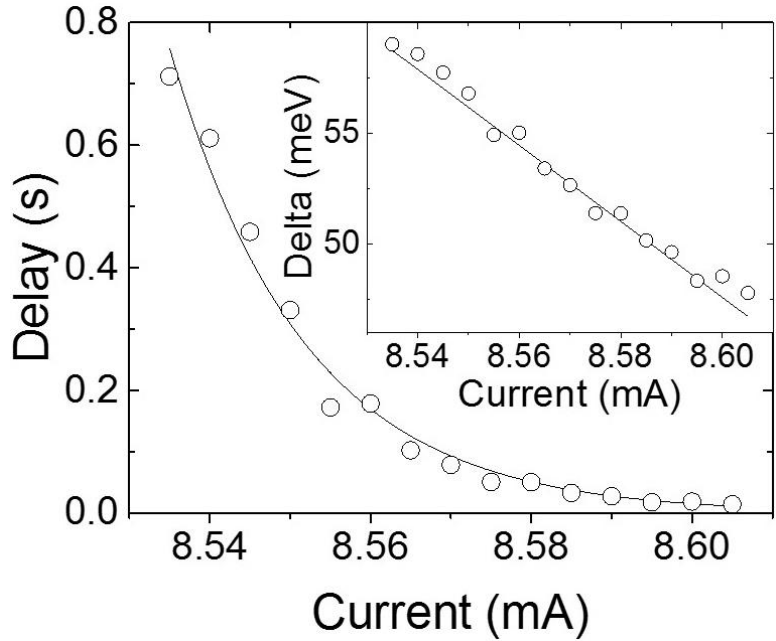

**Figure 3:** (a) Average delay time between application of current step and resistive switching as a function of the final bias measured at 77 K. All switching events were executed from a beginning bias of 8.0 mA. Solid line shows exponential fit. The inset depicts the change in the barrier height, $\Delta$, as a function of the final bias with a linear fit corresponding to a modulation of 172 meV/mA.

Now we can verify this equation by performing our experiment at different temperatures. In our set-up the sample is immersed in liquid nitrogen to maximize the heat removal and minimize any Joule heating from high bias currents. The sample temperature can be accurately controlled in the range from 77-80 K by controlling the pressure (from 750-1020 Torr) in the nitrogen cryostat using an adjustable pressure relief valve. Note that the pressure of the nitrogen column above the sample should also be taken into account in order to determine the actual temperature of the sample using the well-established temperature/pressure relation for liquid nitrogen [5].

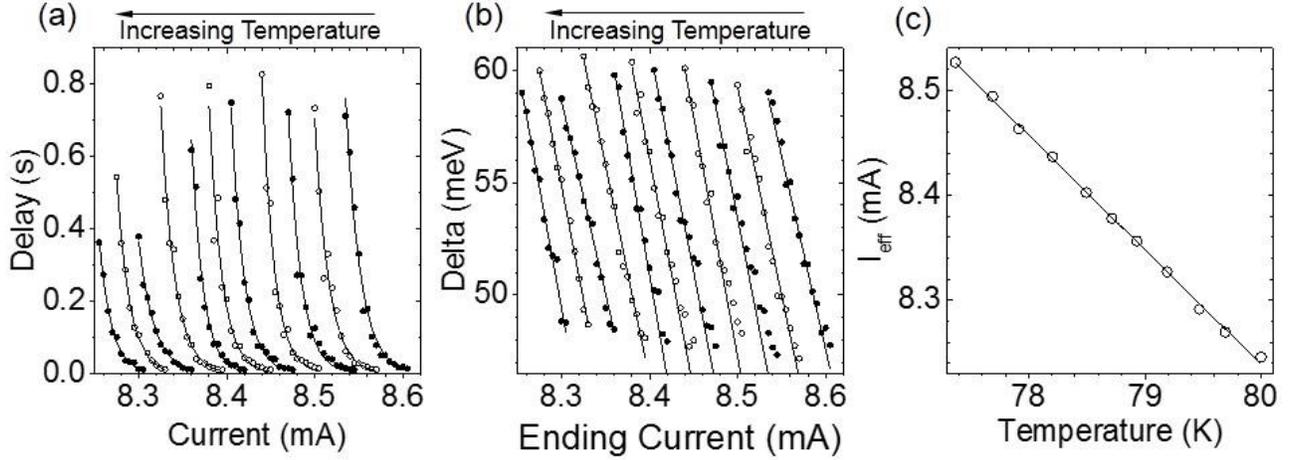

**Figure 4:** (a) Temperature effects of the average delay measurement shown in figure 3. Temperatures range from 77 K at the rightmost curve to 80 K at the leftmost curve. Black and white symbols serve to separate data associated with neighboring temperatures. (b) Delta calculated as a function of temperature ($\tau_0 = 1$ ns). (c) The change in the current, $I_{eff}$ associated with delta equal to 60 meV, as a function of temperature with a linear fit corresponding to a variation of $110 \pm 1$ µA/K.

Figure 4a shows the average switching delay vs applied bias (as in Fig. 3) measured at eleven different temperatures in the range from 77-80 K. Equation 3 was successfully used to fit the data at different temperatures (see solid curves); the curves shift to the left with increasing temperature. Using Eq. 1 we have reconstructed $\Delta(I)$ for different temperatures which are shown in Fig. 4b with the corresponding linear fits. Using the fits for $\Delta$, we can define an effective current, $I_{eff}$, that represents the temperature variation of an average delay time associated with a particular value of $\Delta$. Figure 4c shows $I_{eff}(T)$ for $\Delta = 60$ meV. The temperature dependent variation of $I_{eff}$ is analogous to the previous temperature-dependent measurement of the critical switching current $I_C$. Indeed, the slope of $I_{eff}(T)$ – $110 \pm 1$ µA/K, is comparable to the slope of $I_C(T)$ – $119 \pm 3$ µA/K. We should note that our method to control the temperature may cause some of the observed effects to be associated with the pressure variations. It is possible that the changes we observed in transport properties of our samples are caused directly by the pressure. However, given a very large magnitude of pressure (on the order of GPa) needed to induce transport effects in iridates [6, 7, 8, 9], we estimate the effects from pressure on the order of an atmosphere to be negligible. Therefore we attribute the observed in Fig.4 changes in transport properties to temperature.

In summary, we have investigated the energy landscape of bias induced resistive switching in antiferromagnetic $Sr_3Ir_2O_7$. Using temporally resolved transport measurements, a bias and temperature dependent switching delay time was observed. A model based on thermally assisted transitions over an energy barrier was used to explain the observed behavior. The energy barrier's dependence on applied bias and temperature was quantified. An electric field effect model describing induced structural modifications can be used to explain both discontinuous changes in $Sr_3Ir_2O_7$ resistance and well as previously demonstrated electrically tunable transport in $Sr_3Ir_2O_7$ and $Sr_2IrO_4$ [2,10]. These findings are of high interest for future antiferromagnetic spintronics and spin dynamics with the general goal of high-speed applications controlled by electric fields.

This work was supported in part by C-SPIN, one of six centers of STARnet, a Semiconductor Research Corporation program, sponsored by MARCO and DARPA, by NSF grants DMR-1712101 and DMR-1122603, and by the King Abdullah University of Science and Technology (KAUST) Office of Sponsored Research (OSR) under Award No. OSR-2015-CRG4-2626.